\documentstyle[aps,prl,float,epsfig]{revtex} %galley style

\def\bfomega{\omega\hskip -0.7em\omega\hskip -0.69em\omega}

\begin{document}
\draft
\preprint{HEP/123-qed}

\wideabs{
\title{Rapid dissipation of magnetic fields due to Hall current}
\author{{S.I. Vainshtein$^1$, S. M.
Chitre$^{2}$, and A. V. Olinto$^1$}\\
{$^1$Department of Astronomy \& Astrophysics, The University
of Chicago, Chicago, IL 60637}\\
{$^{2}$Tata Institute of Fundamental Research, Bombay,
India 400 005}}
\date{\today}
\maketitle

\begin{abstract}
{We propose a mechanism for the fast dissipation of magnetic fields which
is effective in a stratified medium where ion motions can be neglected. 
In such a medium, the field is frozen into the electrons  and Hall
currents prevail. Although Hall currents conserve magnetic energy, in the
presence of density gradients, they are able to create current sheets
which can be the sites for efficient dissipation of magnetic fields. 
We recover the frequency, $\omega_{MH}$, for Hall oscillations modified by
the presence of density gradients. We show that these oscillations can
lead to the exchange of energy between different components of the
field. We calculate the time evolution and show that magnetic fields can
dissipate on a timescale of order  $1/\omega_{MH}$. This mechanism can
play an important role for magnetic dissipation in systems with very
steep density gradients where the ions are static such as those found
in the solid crust of neutron stars.}
\end{abstract}
\pacs{PACS number(s): 52.30.Bt, 52.30.-q, 47.65.+a, 97.60.J}
}

\narrowtext
\section{ Introduction}

Rapid dissipation of magnetic fields is currently one of the key problems
in astrophysics. On account of the generally large electrical
conductivities that obtain in astrophysical settings, the Ohmic
dissipation of fields usually takes place on very long time-scales.
However, it is quite often observed that  astrophysical magnetic fields
change topology on a very short time-scale, giving rise to a variety of
transient phenomena. An explanation of such fast changes is crucial to the
understanding of solar activity, in particular,  solar flares
\cite{1},     and other active phenomena observed in stars \cite{1,3}. It
is also known that fast reconnection of magnetic fields is basic to the
operation of non-linear dynamos \cite{4}.

If a current sheet is formed in a plasma, the reconnection takes place
slowly due to the time-scale for the removal of matter from the site of
reconnection, as in the Parker-Sweet mechanism \cite{5}. Indeed, if the
dissipation in the current sheets were to be fast, with the field moving
with Alfv\'en speed toward the current sheet and getting dissipated there
due to reconnection, then the matter that is frozen into the field would
also move with the same speed towards the sheet. The plasma would,
therefore, accumulate at the sheets and halt the reconnection, unless
there is some efficient evacuation process operating at the sheet. The
sheets are usually narrow and the outflow is rather inefficient, even if
it takes place at the Alfv\'en speed \cite{new}. 

We propose a mechanism of fast dissipation
of magnetic fields that occurs at modified Hall frequencies. The mechanism
is relevant in all situation when Hall currents predominate, and
there is a density stratification. In this case, 
the magnetic field follows the (electric) drift velocity of
the electrons. In the presence of density gradients, the profile of
magnetic field changes in such a way that it forms a current sheet.
This steepening of the front is {\it not} accompanied by the flow of
plasma towards the sheet, the drift velocity being parallel to it.
Consequently, the current sheet is formed and  the field is
efficiently dissipated with no accumulation of material in contrast to
the Parker-Sweet reconnection mechanism. 

The modified Hall frequency that we recover below, $\omega_{MH}$, occurs
in a stratified medium when the ions remain static. Two example of such
situations are: penetration waves in low collision plasmas relevant for
plasma switches \cite{shock}, and neutron star crusts. In the case of
penetration waves, the ion response time is long compared to the wave
timescale and the ions are approximately static. In the case of neutron
star crusts, the ions form a very high density lattice of iron rich
nuclei with densities varying from $\sim 10^{6} {\rm g/cm}^{-3}$
to $\sim 10^{11} {\rm g/cm}^{-3}$ under $ 0.8$ km \cite{ST}. % change!
In both
cases, the field dynamic is governed by the  electron drift motion.

In the following sections, we discuss how Hall currents in a stratified
medium can generate fast dissipation in the non-linear regime. In
Section II, we describe the linear Hall oscillations and discuss how
poloidal and toroidal fields exchange energy during these oscillations. 
We also  recover the  modified Hall frequency
for a stratified medium.  We discuss the limitations of the oscillatory
solutions about a stationary configuration in Sec. III. We argue that in
general there is no stationary configuration for large scale magnetic
fields  and that current sheets develop. The oscillations can occur only
``locally", i.e., on small scales. In Sec. IV, we show that the
magnetic field evolution is governed by a non-linear equation similar to
Burgers equation. We solve the evolution for a toroidal field
configuration numerically and show that current sheets develop and
magnetic dissipation is efficient. The dissipation timescale is $\sim
1/\omega_{MH}$. We describe numerical solutions for two configurations: 
a toroidal magnetic field of one polarity; and a toroidal magnetic field
consisting of two oppositely directed fields.  We show that these fields
evolve towards forming  current sheets that rapidly dissipate. In Sec. V,
we relate the dynamics of toroidal fields with that of poloidal fields,
and summarize  the different possibilities of the field evolution.
We close by discussing the application of this physical mechanism
focusing particularly on the  case of neutron stars crusts (Sec.  VI).%change

\section{ Linear oscillations.}

Consider the magnetic field evolution in the case where the motion
of ions can be neglected. This is the case for neutron stars' solid
crust.  We will consider collisional plasma. Then,
the field  evolution follows from Ohm's law,
\begin{equation}
\nabla\times {\bf B}=\frac{4\pi}{c}\sigma({\bf E}+{1\over c} {\bf v}_e
\times{\bf B}),
\label{1}
\end{equation}
where ${\bf v}_e$ is electron velocity. As the conductivity is usually
high, the left-hand side of (\ref{1}) can be neglected, resulting in
\begin{equation}
{\bf E}+{1\over c} {\bf v}_e \times{\bf B}=0,
\label{1a}
\end{equation}
corresponding to the electric drift of electrons.

Taking $\nabla\times$ of equation (\ref{1}), we recover the induction
equation, with the Hall effect,
$$
{\partial {\bf B}\over \partial t}=\nabla\times [{\bf v}_e\times {\bf
B}]-\nabla\times \eta\nabla
\times{\bf B}=
$$
\begin{equation}
-\nabla\times \left( {c\over 4\pi ne}
[\nabla\times {\bf B}]\times {\bf B}\right)-\nabla\times \eta\nabla
\times\bf B,
\label{2}
\end{equation}
where $\eta=c^2/4\pi \sigma$.
Equation (\ref{2}) results in the following energy balance,
\begin{equation}
\frac{1}{2}\frac{\partial}{\partial t}\int B^2 dV=- \int \eta
(\nabla\times {\bf B})^2 dV.
\label{2o}
\end{equation}

In order to describe the evolution of the field in a stratified 
medium, we consider and axisymmetric magnetic field, which can be
expressed as the sum of poloidal and toroidal components,
$$
{\bf B}={\bf B}_p+{\bf B}_t.  $$
In order to simplify the geometry we assume that the radius of the star,
$R$,  is large compared to the wavelengths involved such that we can work  
in Cartesian coordinates on the surface of a sphere. We define the $x$ and
$y$ axes in the horizontal plane as the latitudinal
 and azimuthal (longitudinal) directions respectively, while  $z$ is the
vertical direction. Then, the poloidal field is described by
$$
{\bf B}_p=\{B_x(x,z),0,B_z(x,z)\}, $$
while the toroidal field is given by
$$
{\bf B}_t=\{0,B_y(x,z),0\}. $$
If the resistivity $\eta$ can be neglected, which  is justified in highly
conducting media such as neutron star crusts, then the magnetic energy is
conserved, according to (\ref{2o}). Therefore, all that happens to the
magnetic field are {\it oscillations} about a stationary
configuration. 

Consider, for example, an initial poloidal field, ${\bf B}_0=\{B_0,0,0\}$,
where $B_0$ is a constant background field. Assuming first that the
density is also constant, and considering small perturbations of the
magnetic field of the form,
\begin{equation}
{\bf b}={\tilde{\bf b}}e^{-i\omega t+ik_xx+ik_zz} 
\label{2n0}
\end{equation}
($k_y=0$ because of axial symmetry), we find substituting in \ref{2}
for $\eta \to 0$
\begin{equation}
\omega=\omega_H=\frac{|({\bf k}\cdot \bfomega_e)|c^2k}{\omega_p^2},
\label{2n}
\end{equation}
where $\bfomega_e=e{\bf B}_0/mc$ is the electron cyclotron frequency, and
$\omega_p$ is the plasma frequency. We have thus recovered the well known Hall
oscillations or whistlers. Note that even if  $b_y=0$ initially, i.e., the
toroidal component is absent, it will be generated reaching the level of
the  (perturbed) poloidal component; and thus, the energy will be
exchanged between the poloidal and toroidal components.

The situation is different if the large scale background 
field is toroidal, i.e.,  ${\bf B}_0=\{0,B_0,0\}$. Then a perturbation
of the form (\ref{2n0}) would not result in oscillations (\ref{2n}),
because $({\bf k}\cdot \bfomega_e)=0$. 

Let us now recall that the density
is not a constant, but, rather, it has a steep dependence  on $z$. By
including the spatial dependence of the density in  (\ref{2}), we
recover the Hall frequency modified by the presence of  density gradients,
\begin{equation}
\omega=\omega_{MH}=\frac{({\bf k}\cdot [\bfomega_e\times \nabla n])c^2}
{\omega_p^2n}.
\label{2n1}
\end{equation}
Note that $\bfomega_e$ is a pseudo-vector, and therefore the frequency
$\omega_{MH}$ is a real scalar, as it should be, just as
$\omega_H$ is a real scalar as well, see (\ref{2n}). The phase velocity,
corresponding to (\ref{2n1}) can be written as
\begin{equation}
{\bf v}_{MH}=-\frac{c^2}{\omega_p^2}\left[\frac{\nabla n}{n}\times
\bfomega_e\right]\ ,
\label{2a0}
\end{equation}
and $\omega_{MH} = {\bf k}\cdot {\bf v}_{MH}$.
A similar case is known in low collisional plasmas where the
corresponding wave is called    magnetic penetration wave \cite{shock}. 

The wave described by (\ref{2n1},\ref{2a0}) corresponds to  only
toroidal perturbations due to the chosen initial configurations. In this
special case there is no poloidal field initially and no  
energy exchange occur between toroidal and poloidal components. %change!
However, in  general the two components are present and  this
exchange does take place. In order to see this,  let us return to the
large scale poloidal field, ${\bf B}_0=\{B_0,0,0\}$, taking into account
that the density is a function of
$z$. We look for solutions of the linearized equations in the form,
\begin{equation}
{\bf b}=\{\partial_z a(z),b_y(z),-ik_xa(z)\}e^{-i\omega t+ik_xx},
\label{2a1}
\end{equation}
cf. (\ref{2n0}). Then we obtain the following dispersion relation,   
\begin{equation}
\omega^2a=\frac{({\bf k}\cdot \bfomega_e)^2c^4}{\omega_p^4}(k_x^2-
\partial_z\partial_z)a.
\label{2n2}
\end{equation}
In order to estimate the frequency consider two zones $z_2 \leq z <
z_1$, with density $n_2$, and $z_1 \leq z\leq 0$ ($z=0$ is the top of 
the crust), with density $n_1$, and
$|z_1|=h_1$, and $z_1-z_2=h_2$, $h_{1,2}$ being the scale hight in these
two zones. Assuming that $n_2\gg n_1$, and $h_2\gg h_1$, we obtain 
\begin{equation}
\omega=\frac{\pi}{2}\frac{|({\bf k}\cdot \bfomega_e)|c^2}{\omega_{p_2}^2}
|k_x+1/h_1|,
\label{2n3}
\end{equation}
where $\omega_{p_2}$ is the plasma frequency based on the density $n_2$.
Note that if $k_x \ll 1/h_1$ (large horizontal length scale), the
frequency is essentially the same as $\omega_{MH}$ in (\ref{2n1}).
This is the main characteristic frequency of magnetic
fluctuations in the crust due to the steep density gradient. As seen
from (\ref{2a1}), this mode does involve both poloidal and toroidal
components.

The most general case involves non-linear coupling between the poloidal
and toroidal fields.  Qualitatively the same situation
will take place: if we start with a poloidal field, supported by the
currents in the crust, a toroidal field will be generated. The
current velocity is toroidal, and, according to (\ref{2}), the toroidal
field is stretched out from the poloidal, analogously to the effect of
differential rotation. However, unlike the latter, the Hall current
conserves the energy, and therefore the new toroidal field will grow  at
the expense of the poloidal field. In other words, while the strength of
the toroidal
field is increasing, that of the poloidal component should decrease.%changes 
Of course, the
toroidal field cannot grow indefinitely under these circumstances, and
eventually  the field will either reach some
steady state, or the poloidal and toroidal fields will exchange their
energies, oscillating with frequency $\omega_{MH}$. 

Note, however, that including dissipation may drastically change the
situation, and, in some cases, discussed below in Sec. III, and IV, the
field will rapidly dissipate instead of oscillate. 
\section{The problem of stationary states.}

This simple picture of oscillations implicitly assumes that they proceed
about some stationary state, which presumably exists. 
The large scale background field considered above was uniform and
trivially  stationary. We will show that, in general, the large
scale field is not stationary but evolves with time. 

It is clear  from (\ref{2}) that the stationary state is possible if,
neglecting diffusion,
\begin{equation}
{c\over 4\pi ne} [\nabla\times {\bf B}]\times {\bf B}
=\nabla \Phi,
\label{2a}
\end{equation}
that is, the electric field is  potential.
We will show that condition (\ref{2a}) does not trivially occur even
for extremely
simple topologies, due to the gradient of the density. Indeed,
consider an initial configuration consisting only of a toroidal field.%change

Equation (\ref{2}) for a pure toroidal field can be written as
\begin{equation}
\partial_t B_y+\tilde{v}_x\partial_x B_y+\tilde{v}_z\partial_z
B_y=\eta \nabla^2 B_y,
\label{3}
\end{equation}
where
\begin{equation}
\tilde{v}_x={c\partial_z n\over 4\pi e n^2}B_y-\partial_x\eta, ~~~
\tilde{v}_z=-{c\partial_x n\over 4\pi e n^2}B_y-\partial_z\eta.
\label{3a}
\end{equation}
If we neglect the resistivity in this expression, we
recover the penetration wave velocity (\ref{2a0}), $\tilde{\bf v} \to {\bf
v}_{MH}$ as $\eta \to 0$.

It can be seen from equation (\ref{3a}) that the $x$-component of
the velocity is non-vanishing, due to the vertical gradient of the density.
Note that both the density gradient  and the gradient of the resistivity 
$\eta$, are negligible in the $x$ direction,  and the $\tilde{v}_z$ 
component defined only by the resistivity gradient is also  small.
Since any toroidal field should vanish at least at the two poles,
there is always a latitudinal dependence of the toroidal field, that
is to say that $B_y$ is always a function of $x$. Hence, according to
(\ref{3}), the toroidal magnetic field  can never attain
a stationary state. In other words, the electric field cannot be
irrotational, as in (\ref{2a}), and its non-potential part results in
the time evolution of the magnetic field. 

Note that in infinite space equation (\ref{3}) conserves magnetic flux,
\begin{equation}
\int B_y dxdz ={\rm const},
\label{3b}
\end{equation}
but the  magnetic energy is dissipated according to,
\begin{equation}
\frac{1}{2}\frac{\partial}{\partial t}\int B_y^2 dxdz=-
\int \eta (\nabla B_y)^2 dxdz,
\label{3c}
\end{equation}
which is a particular case of (\ref{2o}).

On the other hand,
for a real toroidal field which should vanish at the poles, i.e.,
at $x=\pm \pi R/2$, the magnetic flux {\it is not conserved}. Indeed,
according to (\ref{3}),
$$
\frac{1}{2}\frac{\partial}{\partial t}\int B_y dxdz=
\int \eta \partial_x B_y(x=\pi R/2)dz 
$$
\begin{equation}
-\int \eta \partial_x B_y(x=-\pi
R/2) dz \ .
\label{3d}
\end{equation}
The right hand side gives considerable contribution when
current sheets are formed at $x=\pm \pi R/2$.

\section{ Toroidal magnetic field evolution: formation of
current sheets.}

\centerline{\bf A. Analytical and Numerical Solutions.}
\medskip

In order to study the evolution of the field, according to (\ref{3}),
we reduce this equation to (neglecting the resistivity gradient,
and resistive diffusion in $z$-direction)
\begin{equation}
\partial_t b +b\partial_x b=\eta \partial_x\partial_x b,
\label{6}
\end{equation}
where
\begin{equation}
b={B_y}{p},~~~~~~~p={c\partial_z n\over 4\pi e n^2}.
\label{6a}
\end{equation}
This is, in fact, the Burgers equation, the exact solution of which is well
known, see, e.g., \cite{14}. First, let us illustrate a solution in the
form of a traveling shock wave,
\begin{equation}
b=b_0\left (1-\tanh\left\{\frac{(x-b_0t) b_0}{2\eta}\right\}\right),
\label{6b}
\end{equation}
where $b_0$ is a constant, cf. e.g. \cite{shock,shock1}. 
The penetration wave (\ref{6b}) does not decay because the magnetic field
is pumped into the system from $-\infty$. Therefore, it is more
appropriate for our purposes to use the general exact solution, which 
we recover by using the transformation
\begin{equation}
b =-2 \eta\partial_x\ln{\xi},
\label{7a}
\end{equation}
to get,
\begin{equation}
\xi =\int_{-\infty}^\infty \frac{1}{(4\pi\eta t)^{1/2} }e^{-(x-x')^2/
(4\eta t)}\xi(t=0,x',z)dx' \ .
\label{8}
\end{equation}
Generally, the toroidal field $B_y$ is a function of both $x$ and $z$,
and, since the $z$-dependence enters only parametrically into 
(\ref{6},\ref{6a}),
the solution (\ref{7a},\ref{8}) can be used for each level $z=const$.

To illustrate the time evolution of the magnetic field, we demonstrate
the following two cases. In the simplest case, we assume that the toroidal
magnetic field does not change sign.
Then the horizontal velocity $\tilde{v}_x$ is expected to drive the field
to
one of the poles, either to the South, or to the North, depending on the
sign of the field. The gradient of the field steepens,   as in a
shock wave, thus forming a current sheet, where the magnetic field
is finally dissipated. In the second case, consider  two toroidal fields
with opposite polarities in the two hemispheres.
The toroidal field vanishes at the equator. We then expect that the 
two toroidal fields can
be driven by the latitudinal velocity $\tilde{v}_x$ towards the
equator, where the current sheet is formed, and the fields are
efficiently destroyed.

The integral in (\ref{8}) was calculated numerically, and then the
distribution of magnetic field $B_y$  was recovered from equations
(\ref{6a}) and (\ref{7a}). Let us discuss the first case where the
toroidal magnetic field does not change sign.  Its
evolution  is depicted in Fig. 1, where the initial field
distribution is indicated by the dashed line. The field profile starts
to steepen in very few turn-over time steps, and moves towards the
polar region. Note that  the magnetic field in Fig. 1 is not 
pumped into the system, cf. (\ref{6b}). Therefore, 
unlike the traveling wave (\ref{6b}), as the magnetic field spreads its
amplitude decreases  keeping the magnetic flux conserved and, thus,
the same area under each curve, see (\ref{3b}). As a result of decreasing
magnetic field, the process slows down, because the penetration velocity
(\ref{2a0}) is proportional to $\bf B$, and therefore it decreases as
well. 
$$
\psfig{file=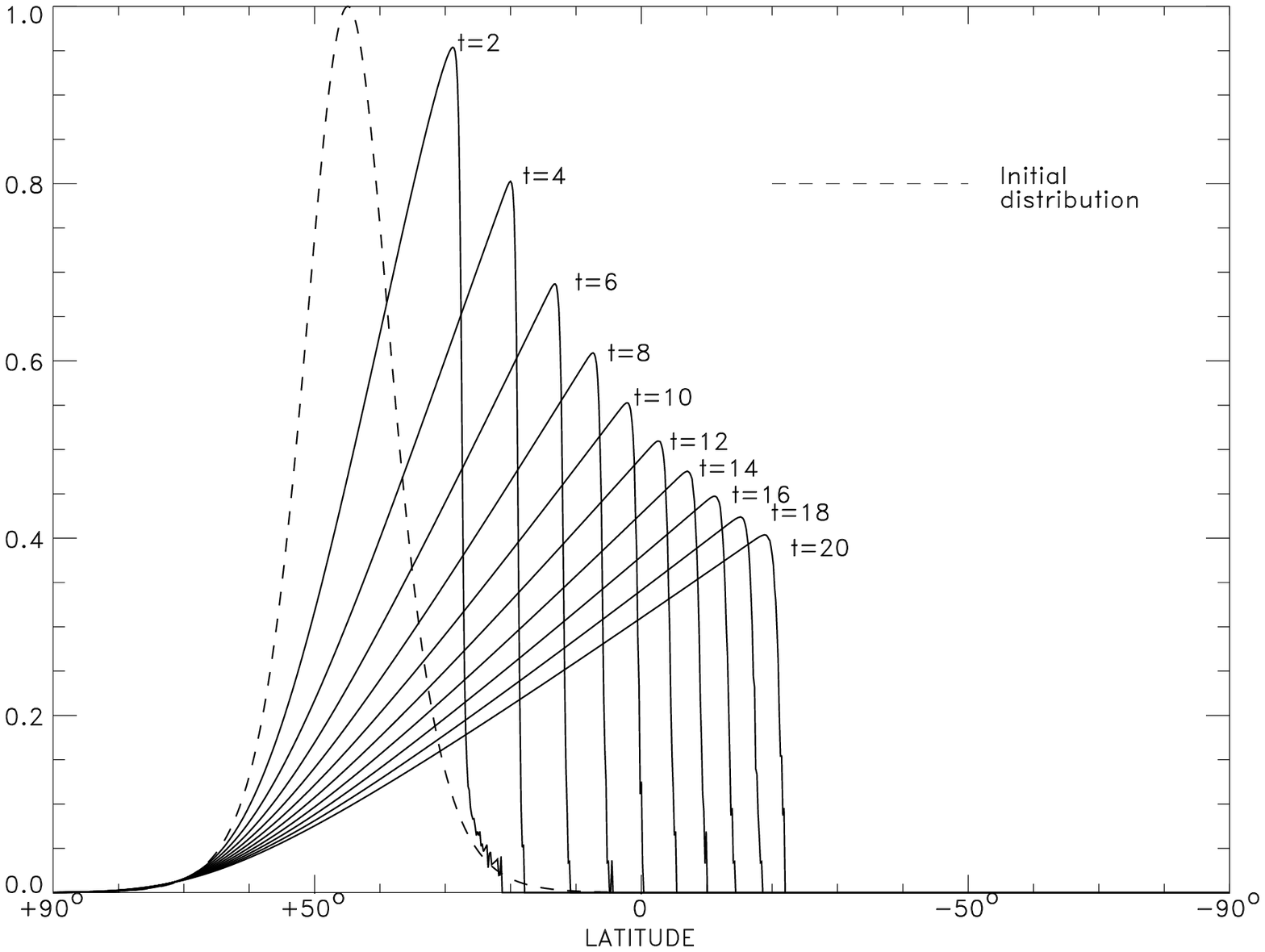,width=3.5in}
$$
\begin{figure}
\caption{Evolution of magnetic field of single polarity in the crustal
region of a neutron star, as it approaches a polar region.
The equator is located at zero latitude, and the
time $t$ is expressed in units of turn-over time $t_0$, given in
(\ref{5a}).}
\end{figure}

In infinite space, both the
shock wave (\ref{6b}) and the solution depicted in Fig. 1 do not result
in dissipation of magnetic field and  the magnetic flux is conserved 
according to (\ref{3b}). The field is only spread out. However, for a
finite case such as that of a star, the boundary conditions at the poles
forces the field to go to zero. 
 When the shock wave reaches the pole, a current sheet is formed and the
field starts to dissipate according to (\ref{3d}). Eventually, the
magnetic flux goes to zero.

In order to see this dissipation at a zero-point, we proceed to the second 
example. Namely, consider the toroidal field changing sign at $x=0$. 
It
is straightforward to construct a solution, analogous to the traveling wave
(\ref{6b}), 
\begin{equation}
b=-b_0\tanh\left\{\frac{x b_0}{2\eta}\right\}.
\label{8c}
\end{equation}
Similarly to the Parker-Sweet solution \cite{5}, the magnetic field of
opposite polarities is transported from $x\to \pm\infty$ with ``velocity"
$\pm b_0$, and is dissipated
at $x=0$. The solution is stationary because the boundary
conditions are: $B_y(x\to \pm\infty)\to \pm b_0/p$, see (\ref{6a}). 
$$
\psfig{file=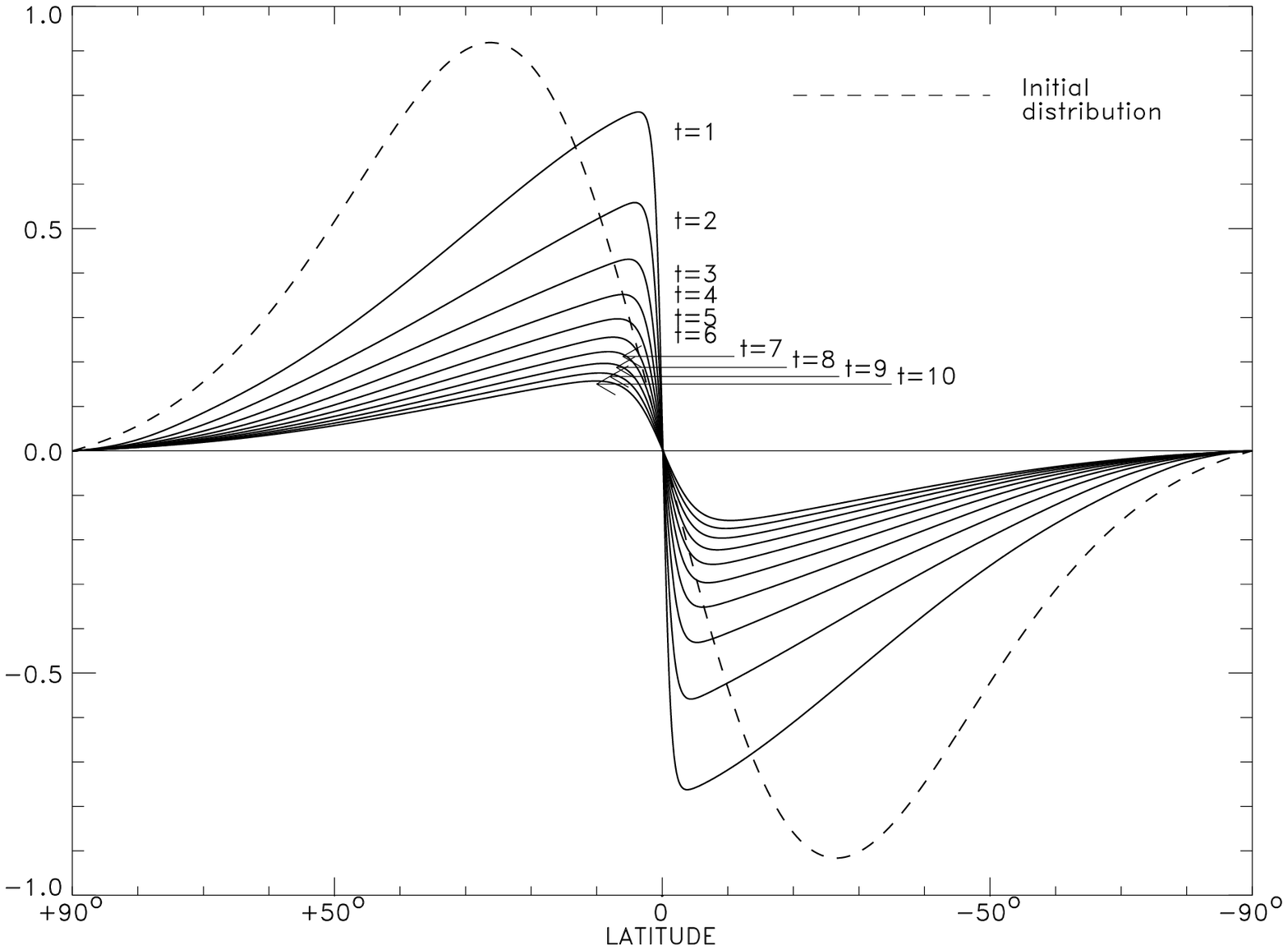,width=3.5in}
$$
\begin{figure}
\caption{The evolution of magnetic fields of opposite polarities
in the two hemispheres. In this case, the two fields are approaching
each other to form a current sheet at the equator, where the magnetic
energy is efficiently dissipated.}
\end{figure}
In Fig. 2 we display the evolution of a toroidal field with
opposite polarities in each hemispheres and  $B_y(x=0)=0$. The field also
vanishes at the poles,
$B_y(x=\pm \pi R/2)=0$, which makes the solution  
evolve in time (again, in contrast to the infinite space case). Here,
the two fields get compressed into each other and form a sharp
gradient of magnetic field at the equator. Note that the total magnetic
flux is zero and, of course, trivially conserved. As to the magnetic
energy, it decreases dramatically because of the very efficient Ohmic
dissipation at the equatorial region. In this region, a current sheet is
formed  in practically only one turn-over time and the field 
dissipates in the same time scale. An analytical estimate of the
dissipation time is given in the next subsection; it illustrates why the
dissipation observed numerically is so efficient.

\medskip
\centerline{\bf B. Physical Interpretation of the Mechanism.}
\medskip

In order to develop a physical interpretation of the solutions above, we
draw a few analogies.  The equation for the magnetic field
(\ref{2}) resembles the  vorticity equation for incompressible
hydrodynamics with high Reynolds numbers. Therefore, the
modified Hall drift should lead to a situation analogous to a magnetic
turbulent state
\cite{9}. Another interpretation of our solutions is the
nonlinear interaction of different wavenumber Hall oscillations resulting
in an energy cascade to small scales
\cite{9a}. As a result, the field gradients steepen and we get an enhanced
local rate of Ohmic dissipation which  provides an effective
mechanism for the dissipation of magnetic energy.

These analogies, although helpful, cannot be taken completely due to the
topological constraints on magnetic fields. For instance, since 
the magnetic structures are frozen into the  electron fluid, they are
generally more persistent than vortices in hydrodynamics. The 
topology of magnetic field cannot be easily changed, a situation similar%change
to what one obtains in MHD  \cite{4,10}.
The magnetic structures we considered are non-stationary due to global
effects (like the boundary conditions at the poles) and not  locally
unstable like the case of vorticity and magnetic turbulence. Besides,
in the case
of magnetic turbulence,  the characteristic frequency coincides with the 
Hall oscillations frequency (\ref{2n}), which is smaller than
$\omega_{MH}$ from (\ref{2n1}) for situations with sharp density
gradients. Therefore, our mechanism is more efficient.

$$
\psfig{file=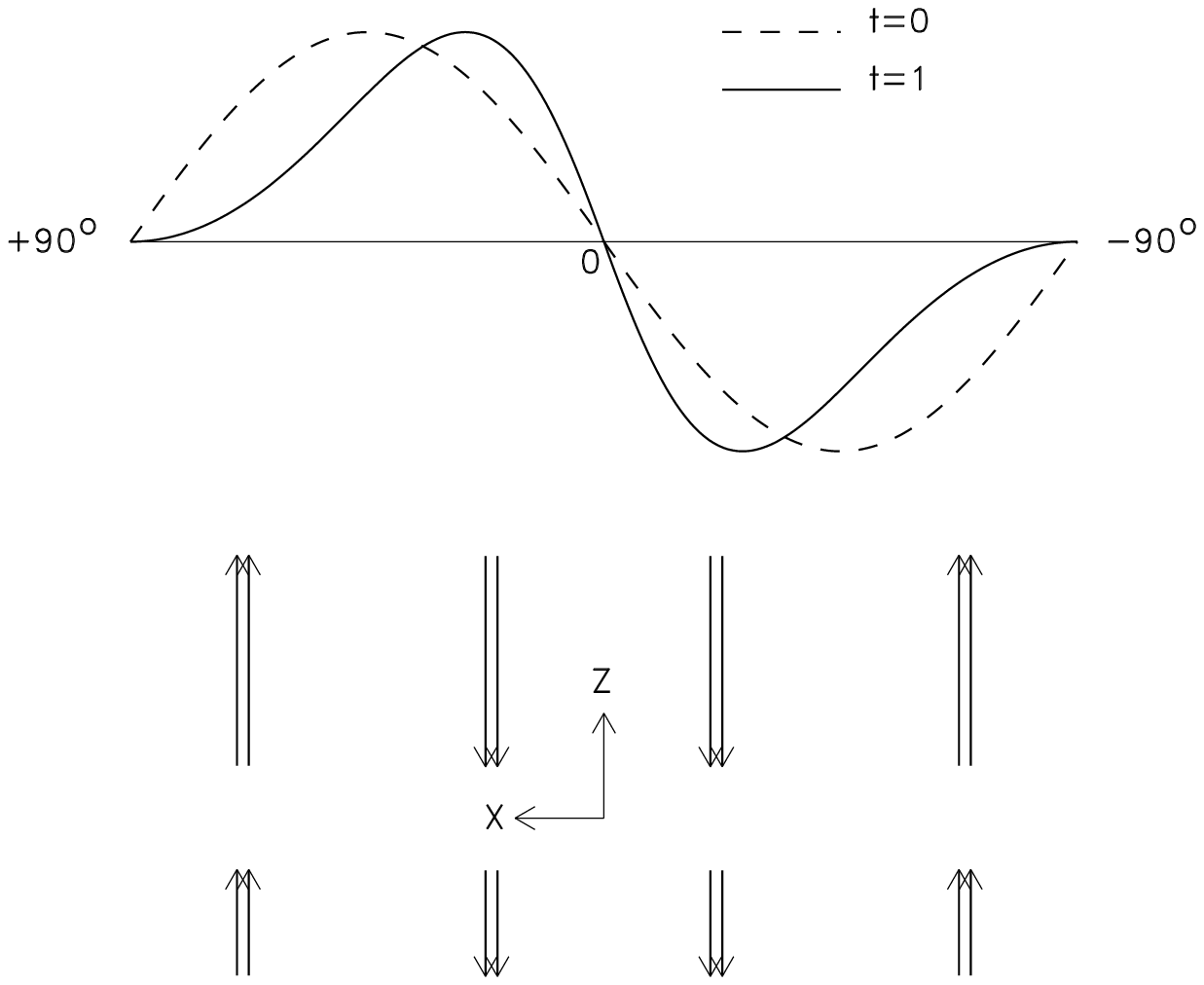,width=3.5in}
$$
\begin{figure}
\caption{Origin of the penetration velocity.     
The magnetic field evolution is depicted on the top of the figure. The
direction of the field is out of the page.  It can be seen that the
field profile steepens during the evolution due to the change in drift
velocity (double arrows).  The drift proceeds in the vertical direction,
and, therefore, there is no accumulation of matter at the current
sheet.  }
\end{figure}
In magnetohydrodynamics,   fast reconnection  encounters difficulties
\cite{1} 
because the rapid transport of magnetic fields towards the current sheet,
where the energy
is dissipated, is accompanied by a plasma movement in the same direction.
The evacuation of matter from the current sheet limits the rate of
reconnection: as matter accumulates and  the pressure  increases
eventually halting the movement of the magnetic field towards the
current sheet, thus preventing further reconnection. The speed of the
evacuation is limited by the Alfv\'en velocity in a narrow current sheet.
In our mechanism, we do not encounter this  difficulty, because
$\tilde{v}_x$ does not transport the mass.
Indeed, according to (\ref{6},\ref{6a}), there is no outflow from the
current sheet: the magnetic field is transported only to the sheet by
the penetration velocity $\tilde{v}_x$.  This is evident from the %change
exact  solution (\ref{8c}).

In order to understand why the modified Hall drift or  penetration
velocity does not transport any mass, we first note 
that, generally,  the penetration velocity (\ref{2a0}) is  {\it not
parallel} to the electric drift velocity (\ref{1a}).  So the question
arises,  why the magnetic field is moving in the direction of the
penetration velocity in the first place? The answer is  illustrated in
Fig. 3. For simplicity, we choose to illustrate a field that
 depends only  on the $x$-coordinate  (as in (\ref{8c})). 
The electric drift of electrons (\ref{1a})  can be easily found from 
Amper's law,
\begin{equation}
{\bf v}_e=-\frac{c}{4\pi ne}\nabla {\bf\times B},
\label{5oo}
\end{equation}
and, clearly, it proceeds in the {\it  vertical} direction (depicted
by double arrows in Fig. 3). It follows from (\ref{5oo}) that
\begin{equation}
\nabla\cdot n{\bf v}_e=0.
\label{5oo1}
\end{equation}
Due to the density gradient, the plasma gets compressed as it moves down
and the descending motion  decelerates, as follows from (\ref{5oo1}). As
a result, the magnetic field amplitude increases. On the other hand, the
ascending motion is accompanied by the decompression of plasma, and
correspondingly,  the field amplitude decreases. As a
result, the field profile steepens, as if there were a motion of plasma
toward the current sheet, that is in the {\it horizontal} direction.
However, as mentioned, the real drift motion proceeds parallel to the
sheet (in the vertical direction), and therefore there is no accumulation
of matter in the sheet. These circumstances make the fast
dissipation of magnetic field possible.

As mentioned above, the penetration wave (\ref{2n1},\ref{2a0}) is known 
to propagate as a shock wave in low collisional plasma \cite{shock,shock1}. 
Due to the conservation of the magnetic flux, (\ref{3b}), either the
magnetic field is only transported by the shock wave (\ref{6b}) or it 
just  disperses to infinity. Only the presence of zero-points result in
the destruction of magnetic flux. This happens either at the poles or
between two shock waves with opposite magnetic fields that collide. The
collisional front width of the shock wave coincides with current sheet
thickness. Indeed,  the balance 
between the convective and resistive terms in equation  (\ref{3})
appears in a current sheet of thickness, 
\begin{equation}
\delta=\frac{\eta}{\tilde{v}_x},
\label{5b}
\end{equation}
coinciding with characteristic length of both (\ref{6b}) and
(\ref{8c}). Recall that the penetration velocity $\tilde{v}_x$ is
defined in (\ref{3a}), and, in a more specific way, in (\ref{2a0}).

It is known that the time scale of magnetic dissipation, $t_0$, is
entirely defined by the velocity with which the magnetic field is moving
to the current sheet \cite{new}. As seen from the exact stationary
solution (\ref{8c}), this speed is in effect the penetration velocity 
$\tilde{v}_x$. Therefore,
\begin{equation}
t_0=L/\tilde{v}_x,
\label{5a}
\end{equation}
$L$ being the macroscopic latitudinal scale. It is useful to confirm
this dissipation rate estimate analyzing the energy dissipation directly 
from (\ref{3c}).
Namely, we estimate $\nabla B_y$ in (\ref{3c}) as
$B_y/\delta$, and the area occupied by the current sheet is
$S_\delta=\delta L$.
Then, we get from (\ref{3c}),
\begin{equation}
\frac{B_y^2}{t_0}L^2\approx \eta \frac{B_y^2}{\delta^2}S_\delta, 
\label{rate}
\end{equation}
from which $t_0$ is recovered as in (\ref{5a}). 
Note that the
dissipation time (\ref{5a}) is independent of the resistivity
$\eta$, and therefore the process considered here is fast. This explains
why the dissipation is so efficient in the numerical results shown    
in Fig. 2.

We finally note that the efficient dissipation depicted in Fig. 2
proceeds when the penetration velocities of the two toroidal fields 
point to each other, and therefore they collide. If we change sign of
magnetic fields, then, according to (\ref{2a0}), the penetration
velocity changes its direction, and, as a result, the two toroidal
fields would not collide, but instead drift to the polar regions, where
they will eventually decay. As mentioned above in Sec. IVA, the latter 
process is much less efficient
because the magnetic field strength decreases as the fields move to the
poles, as seen from Fig. 1, and therefore the penetration velocity
decreases. 

\section{ Evolution of the poloidal field.}

In Figs. 1 and 2, we have shown the solution of our numerical calculations%change
for the evolution of toroidal fields with different initial profiles.
Since the toroidal and poloidal
fields are coupled through non-linear oscillations, we expect the
dissipation of toroidal fields to cause the eventual decay of the
poloidal field. The exact evolution of the poloidal field is a %change
harder problem to solve
at this stage and we leave it for future studies. Below, we only discuss
the expected qualitative behavior of the poloidal field.

As we saw in Sec. II, the linear
oscillations exchange energy between the poloidal and toroidal
components. In other words, an initial poloidal field would generate a
toroidal one.  The generated toroidal field could have the
nonlinear evolution depicted on either Fig. 1 or Fig. 2. 
In the case that the generated toroidal field is in the configuration of
Fig. 1, it would slowly drift to the poles. 
We expect that, in this case, one would observe
oscillations, because the dissipation is inefficient. 

Consider now the case when the toroidal fields are generated
with the configuration as in Fig. 2. In order to
follow this generation in the nonlinear case, we write, according to
(\ref{2}),
\begin{equation}
\frac{\partial}{\partial t}B_y= -\frac{\partial}{\partial z} \left({c
\over 4\pi ne}j_yB_z\right)-
\frac{\partial}{\partial x}\left({c\over 4\pi ne} j_y B_x\right),
\label{9}
\end{equation}
where
\begin{equation}
j_y=\frac{\partial}{\partial z}B_x-\frac{\partial}{\partial x}B_z,
\label{10}
\end{equation}
is the azimuthal current. Note that, for neutron star crusts, the density
gradient in the radial direction is extremely
steep in the crustal regions, spanning some nine orders of magnitude
over a distance of a few hundred meters below the surface.
The variations of  other quantities in equation (\ref{9}) can
therefore be  neglected to get
\begin{equation}
\frac{\partial}{\partial t}B_y=-j_yB_z \frac{\partial}{\partial z} {c
\over 4\pi ne}= \frac{c\partial_z n}{4\pi en^2}j_yB_z.
\label{11}
\end{equation}
It is evident from (\ref{11}) that the toroidal field can always be
generated due to the sharp density gradient, unless the poloidal field
in the crustal region is current free. If, indeed, the field is anchored
in the core (meaning that the currents supporting the field are
confined there), then there is no Hall current present. At present, the
locus where the field is anchored in neutron stars is a matter of debate
with no clear resolution (see, e.g.,\cite{ST,6}). 

If we assume that part of
the current supporting the field is present in the crustal layers, then
the toroidal field will be generated from the poloidal field by
the process outlined above in (\ref{11}). On the other hand,
the total magnetic energy is essentially conserved (apart from weak
Joule dissipation),
which means that the newly generated toroidal field would result in
a back reaction on the poloidal field in such a way that the energy of the
latter is decreased. In the absence of Ohmic dissipation, the toroidal
field would grow to a certain level, and then start to decrease, thus
presenting an oscillatory behavior, as described earlier at the end of
Sec. II.
If we incorporate Ohmic diffusion in the nonlinear case the both  
the  toroidal and the poloidal fields should eventually decay.

Again, in the case of Fig. 1, the toroidal fields would slowly drift to
the poles and  we expect to observe oscillations, because the dissipation
is inefficient. In the case of Fig. 2, the toroidal field is efficiently
dissipated and consequently, according to equation
(\ref{2o}), both the poloidal and toroidal fields decay.
Indeed, the Ohmic dissipation is now increased  due to the presence
of current sheets, so that the equation (\ref{2o}) can be written in the
form
\begin{equation}
\frac{1}{2}\frac{\partial}{\partial t}\int B^2 dV=- \frac{1}{t_0}\int
B_y^2 dV.
\label{11a}
\end{equation}

In order to follow the evolution   of the poloidal field,
we introduce efficient dissipation discussed above  into equation (\ref{11}),
to get
\begin{equation}
\frac{\partial}{\partial t}B_y=
\omega_{MH} B_p  -\frac{B_y} {t_0},
\label{12}
\end{equation}
and, because the Hall current conserves the total energy, $\int(B_p^2
+B_y^2)dV$, the
back reaction of the toroidal magnetic field on the poloidal
component can be expressed analogously,
\begin{equation}
\frac{\partial}{\partial t}{B}_p= -
\omega_{MH} {B}_y.
\label{13}
\end{equation}
Seeking solutions $\sim e^{\gamma t}$, we find a dispersion relation,
\begin{equation}
\gamma= -\frac{1}{2t_0}\pm \sqrt{\left(\frac{1}{2t_0}\right)^2-
\omega_{MH}^2}.
\label{14}
\end{equation}
It can be seen from
equation (\ref{14}) that the decay time for the poloidal component
is also of the order of $t_0$.

To summarize, we can delineate three regimes for magnetic fields in the
crusts of neutron stars: %change on previous line

(1) The currents supporting the fields in the crust are anchored in the
core, i.e., no currents in the crust. Then, there is no Hall current (by
definition), and no evolution of the fields related to the processes we
described here. 

(2) The currents or part of the current are situated in the crust. Then,
the poloidal field inevitably generates toroidal fields, which,
depending on the sign of the initial poloidal field, may result in either
penetration velocity pushing these toroidal fields apart or pushing
them together. In the first case, the dissipation is less
efficient, being limited to slow decay at the poles (Fig. 1). Although 
slower than the equatorial case, the decay at the poles is still 
faster than the general Ohmic decay.

(3) In the second case, when the toroidal fields are pushed together as
in Fig. 2, we expect both toroidal and poloidal fields to decay according
to  (\ref{14}).

Note that in any of the above cases there would be oscillations on
scales small compared with the radius of the star (i.e., in the geometric
optics limit), with frequency $\omega_{MH}$. These oscillations in
general will decay with  Ohmic decay time. 

\section{Discussion}

Magnetic fields are an important feature of neutron
stars since, together with the rapid rotation of the star, they
determine  the characteristics of the pulsar emission. The source of a wide
%change above
range of magnetic field strengths ($\sim 10^8 -10^{15}$G) associated with
neutron stars is yet to be well understood. The seven orders of
magnitude span may be attributed to the different environments in which
neutron stars are present: from isolated objects to accriting members
of binary systems. This range could also  be the result of different
%change above
conditions at the time of birth of neutron stars, such as the
gravitational collapse of the progenitor massive star or the
accretion-induced collapse of a white dwarf \cite{6}.  In any case, the
very high electrical conductivity renders the Ohmic decay inefficient
with typical accreting of the order of billions of years. If neutron
star fields decay over their observable lifetime, an alternative decay
mechanism is necessary to explain this behavior.

One of the uncertainties concerning the evolution of neutron star
magnetic fields is their location in the stellar interior. Should the
field be a fossil remnant left over from the progenitor star, it
could permeate the whole body of the neutron star. On the other hand,
if the magnetic field is generated after a neutron star is born
via a battery effect  or  a dynamo process \cite{7}, it is likely to be
confined to its  crustal layers. As we discussed in the previous section,
the exact location of the currents will determine if the mechanism
proposed here is operating in neutron stars or not.

For instance, the interaction between differential rotation and
magnetic field during  the first few seconds of a nascent neutron star's life
%change above
would generate strong toroidal magnetic fields in the
subsurface layers of the star. 
%change below
With the rapid cooling of the star, the crust solidifies with the ions
forming a lattice in the presence of relativistic electrons. Some
fraction of the toroidal field will have different signs in the Northern
and Southern hemispheres, like the one illustrated in Fig. 2.
%end of changes
Under these conditions the magnetic field is frozen into the
electron gas  and Hall currents in the crustal layers can arise and our
mechanism will be effective. In contrast, the Ohmic dissipation in the
crustal layers  takes place on a very long time scale.

If part of the currents %change
supporting the fields are situated in the crust, we can use our
mechanism to estimate the timescale for rapid dissipation to occur.
Taking typical numbers for the crustal layers of a neutron star,
at density scale height $h$ of $10^4 cm$, 
$n=10^{34}cm^{-3}$  and the  magnetic field is $10^{12}G$, then
$\tilde{v}_x\approx  10^{-8}cm/sec$. The corresponding time scale
$t_0=L/\tilde{v}_x$, where $L$ is horizontal scale of the magnetic field,
is  $t_0=10^{14} sec\approx 3$ million
years, assuming $L=10^6 cm$. On the other hand, for a density scale height of
%change above
$3\cdot 10^3 cm$, then $n=10^{32}cm^{-3}$, we have for $B=10^{12}$G,%change
$\tilde{v}_x\approx  10^{-5} cm/sec$,  therefore,
$t_0=10^{11} sec\approx 3000$ years. Finally, if we take a scale height of
$10^3 cm$, then the density is $n=10^{30}$, and $\tilde{v}_x\approx
10^{-3} cm/sec$, and $t_0=30$ years. In a real neutron star,
all of these time scales are present if currents occur throughout the
crust. 

Indeed, due to the sharp gradient of
electrical conductivity in the crustal region, we can consider the depth
$10^3 cm$ as a boundary between two layers, with different $\sigma_{1,2}$,
index $1$ corresponding to the upper layer, and index $2$  to the lower
layer, %change
with $\sigma_2\gg \sigma_1$. Then, the tangential component of the electric
%important change above
field is continuous, resulting in \cite{15}
\begin{equation}
\frac{(\nabla\times {\bf B})_1}{\sigma_1}=
\frac{(\nabla\times {\bf B})_2}{\sigma_2}.
\label{8a}
\end{equation}
It follows from (\ref{8a}) that the currents are much
stronger in the inner layer in a quasi-steady state. Another way to see that
the currents are pumped down the crustal area, is directly from (\ref{3a}):
$\tilde{v}_x =-\partial_x\eta\sim \partial_x\sigma/\sigma^2$.  As the
conductivity  increases inwards, this part of velocity results in pushing
down the magnetic flux. Therefore, if
the initial currents are evenly distributed in the crustal area, the
upper currents dissipate in short time scale (30
years),  currents in  deeper layers  
dissipate over longer time scales ($\sim $ 3000 years), while the whole
crustal field lasts for a few million years.

As we mentioned before, it is not known which part of the currents
supporting the poloidal field is situated in the crust \cite{6,16}.
In any event, that part of the crustal currents can dissipate via our
mechanism in a very short time scale while the field anchored in the core
may remain for time scale comparable with the age of the universe. %change
It is possible that pulsars with relatively low observed
magnetic fields indicate a core component of $\sim 10^8$ G, while pulsars with
fields of order $10^{12}$ G are younger and have not had time to lose their
crustal field component. As isolated pulsars lose their crustal magnetic
field due to rapid decay, they also slow down, in the process crossing the
death line to become unobservable. We  suggest that as neutron stars
in binary systems lose their crustal magnetic fields, they permit an
increased rate of accretion that spins them up to give rise to the
millisecond pulsar population.

{\bf Acknowledgments} We thank P. Goldreich for stimulating our interest
in these problems and for numerous encouraging and valuable discussions.
We also thank R. Rosner, R. Z. Sagdeev, L. I. Rudakov, J. Drake, E. N.
Parker, D. Lamb, V. Krishan,  D. Bhattacharya, and R. Epstein for 
useful comments. One of us (SMC) is grateful to Ed van den Heuvel for 
hospitality at the Astronomical Institute of the University of 
Amsterdam which provided the opportunity to study the role of 
crustal gradients of density and electrical conductivity. The research 
was partly supported by NSF through the collaborative US-India project 
No. INT-9605235, NSF grant AST 94-20759, and DOE grant 
DE-FG02-90ER40606 at the  University of Chicago.

\end{document}